\documentclass[runningheads,a4paper]{llncs}

\usepackage{amssymb}
\usepackage{graphicx}
\usepackage{verbatim}
\usepackage{amsmath}
\usepackage[misc]{ifsym}

\usepackage{url}
\urldef{\mailsa}\path|{maxuecheng|
\urldef{\mailsb}\path||
\urldef{\mailsc}\path|}@iie.ac.cn|
\newcommand{\keywords}[1]{\par\addvspace\baselineskip
\noindent\keywordname\enspace\ignorespaces#1}

\newtheorem{thm}{Theorem}

\newtheorem{lem}{Lemma}

\newtheorem{defn}{Definition}
\newtheorem{rem}{Remark}

\author{Xuecheng Ma\inst{1,2}\and Xin Wang\inst{1,2}\and Dongdai Lin\inst{1}}

\begin{document}

\mainmatter  

\title{Anonymous Identity-Based Encryption with Identity Recovery\thanks{A preliminary version of this work will appear in the proceedings of ACISP 2018}}

\titlerunning{Anonymous Identity-Based Encryption with Identity Recovery}

%
%
%

\authorrunning{X.Ma et al.}



\institute{State Key Laboratory of Information Security, Institute of Information Engineering, Chinese Academy of Sciences, Beijing 100093, China \and School of Cyber Security, University of Chinese Academy of Sciences, Beijing 100049, China \\
\mailsa
\mailsb
\mailsc\\}
%
%

\maketitle

\begin{abstract}
Anonymous Identity-Based Encryption can protect privacy of the receiver. However, there are some situations that we need to recover the identity of the receiver, for example a dispute occurs or the privacy mechanism is abused.  In this paper, we propose a new concept, referred to as Anonymous Identity-Based Encryption with Identity Recovery(AIBEIR), which is an anonymous IBE with identity recovery property. There is a party called the Identity Recovery Manager(IRM) who has a secret key to recover the identity from the ciphertext in our scheme. We construct it with an anonymous IBE and a special IBE which we call it testable IBE. In order to ensure the semantic security in the case where the identity recovery manager is an adversary, we define a stronger semantic security model in which the adversary is given the secret key of the identity recovery manager. To our knowledge, we propose the first AIBEIR scheme and prove the security in our defined model.
\keywords{IBE, anonymous, identity recovery, testable}
\end{abstract}

\section{Introduction}
Public key encryption is one of the most important primitives in cryptography, which was presented in the great paper  titled ``New Directions in Cryptograph'' in 1976 \cite{diffie1976new}. Public key encryption solves the problem that the sender and the receiver should share a common secret key which is not known to the adversary before communicating. One of the disadvantages in public key encryption is using certificate to bind the public key to the identity of its owner. The issue of management of certificates is complex and cumbersome.

In 1984, Shamir \cite{Shamir84} introduced the concept of Identity-Based Encryption (IBE) which solved the problem. IBE is a generalization of public key encryption where the public key of a user can be arbitrary string such as an e-mail address. The first realizations of IBE are given by \cite{SOK00,BF01} using groups equipped with bilinear maps. Since then, realizations from bilinear maps \cite{DBLP:conf/eurocrypt/BonehB04a,DBLP:conf/crypto/BonehB04,Waters05,Gentry06,Waters09}, from quadratic residues modulo composite \cite{Cocks01,BGH07}, from lattices \cite{GPV08,CHKP10,ABB10a,Boyen10} and from the computational Diffie-Hellman assumption \cite{DG17a} have been proposed.

In order to protect the privacy of the receiver, Boyen \cite{Boyen03} first explicitly stated the concept of anonymous IBE\footnote{In fact, Boyen gave an identity-based signcryption with a formalization of sender and recipient anonymity}, where the ciphertext does not leak the identity of the recipient. In fact, \cite{BF01} is the first anonymous IBE scheme although they did not state it explicitly. Since then, there are some follow-up works realized from bilinear maps \cite{BW06a}, from quadratic residues modulo composite \cite{DBLP:conf/ctrsa/AtenieseG09}, from lattices \cite{GPV08,ABB10a} and from the computational Diffie-Hellman assumption\cite{BLSV17}.

Anonymous IBE protects the privacy of the message and the receiver's identity in the meantime, but we can only recover the message. However, there are some situations where we need to recover the identity of the receiver, for example a dispute occurs or the privacy mechanism is abused. In a mail system, there is a need to keep the receiver anonymous for everyone except the mail sever who will forward the mail to the receiver. \emph{So can we extract the identity from an anonymous IBE ciphertext with some secret information}? In this paper, we present a new primitive called \textit{anonymous identity-based encryption with identity recovery}(AIBEIR) which can solve this problem. AIBEIR is a special anonymous IBE which has an additional property that the identity recovery manager can recover the identity with a secret key. But the identity recovery manager can not get any information of the message from the ciphertext. Formally, AIBEIR is semantic secure even when the identity recovery manager is the adversary.

\subsection{Motivations}
On the one hand, anonymity protects user's privacy. On the other hand, unconditional privacy may lose supervision and cause illegal behavior. To balance the anonymity and accountability in anonymous IBE schemes, PKG can send the recovery key to a manager who takes charge of recovering identities of suspected ciphertexts. A similar notion was presented in group signature where a group manager can reveal the member identity registered to is group. But there is a subtlety that signature can be verified while general encryption does not have verifiable property. If a mole want to communicate with his partners, he will choose other secure encryption schemes other than encryption schemes with recovery property. In fact, it is inefficient if every ciphertext should be verified before delivering. So we can just restrict that the anonymous IBE with recovery is the only choice. For example, in the army, the internal communication tool is deployed with anonymous IBE with recovery. If some ciphertext is suspected to contain sensible data which is not allowed to send to the recipient.

In fact, anonymous IBE with recovery identity subdivides the privacy which makes the ciphertext anonymous for all users except some privileged supervisors. Imagine that how can an anonymous IBE encryption under the destination IP address be transferred. Routers with recovery key can recover the corresponding identities of the ciphertext which makes it viable. Similarly, in a mail system deployed with anonymous IBE, the server does not know which one is the recipient.

\subsection{Our Contributions}

We propose a new cryptographic primitive called \textit{anonymous IBE with identity recovery}. We first define the model and security notions of AIBEIR. We then present a method to convert an anonymous IBE into AIBEIR with the help of testable IBE and prove that the new scheme satisfies the security we defined. A testable IBE is an IBE which can test whether ciphertext $c$ is a ciphertext under identity $id$ given $c$ and $id$. It is obvious that a testable IBE is not anonymous. We will show that \cite{DBLP:conf/eurocrypt/BonehB04a,Waters05} and their variations are testable IBEs. AIBEIR consists of four parties, a Private Key Generator(PKG), an Identity Recovery Manager(IRM), a sender, and a receiver. There are five procedures in an AIBEIR scheme. They are setup procedure, extract procedure, encrypt procedure, decrypt procedure and recover procedure.

Besides correctness and anonymity, we introduce two new security notions in AIBEIR. The first is a stronger semantic security, where the identity recovery manager is the adversary. The second is recovery, which ensures that the recovery is reliable and no adversary can fool the identity recovery manager. Finally, We prove the security of our AIBEIR scheme according to our security notions. To the best of our knowledge, our construction is the first anonymous IBE scheme with the identity recovery property.

To construct an AIBEIR scheme, we first encrypt the plaintext by a testable IBE and encrypt the testable IBE ciphertext using an anonymous IBE. Moreover, we encrypt the receiver's identity under the recovery manager's identity. The anonymity is guaranteed by the anonymous IBE and the stronger CPA security is guaranteed by the security of the testable IBE. Given the master secret key of the anonymous IBE, identity recovery manager obtains the identity and the testable IBE ciphertext by decrypting corresponding ciphertext, respectively. Then, check whether the testable IBE ciphertext is under the identity and output the identity if the test algorithm outputs 1.
\subsection{Related Work}

Identity-based cryptosystems were introduced by Shamir \cite{Shamir84}. The first realizations of IBE were given by Boneh, Franklin\cite{BF01} and  Sakai \textit{et al} \cite{SOK00}. Boneh and Franklin gave the security model and their proposal is the first anonymous IBE. The anonymity was first noticed by Boyen \cite{Boyen03}. Another view of Anonymous IBE is as a combination of identity-based encryption with the property of key privacy, which was introduced by Bellare \textit{et al}\cite{BellareBDP01}. A similar concept called Identity-Based Group Encryption(IBGE) was presented by Xiling \textit{et al} \cite{LuoRLHLWXW16}. Traceability in their scheme is similar to recovery in ours. But there are some differences between IBGE and AIBEIR. On the one hand, we do not have \textsf{Verify} algorithm which is used to verify whether the ciphertext belongs to the group. On the other hand, our construction is implemented by IBEs while they utilized PKE, IBE and ZKP(\emph{Zero-Knowledge Proofs}) to construct their scheme. We do not think their scheme is a ``pure'' IBE because of the use of PKE. Recently, \cite{Gupta} pointed that the zero-knowledge proof used in \cite{LuoRLHLWXW16} leaks much more information, due to which the verifier who is honest but curious will be able to identify the designated recipient. They proposed a construction with six random oracles.
\subsection{Organization}
This paper is organized as follows. In Sect. 2, we introduce definitions that we use throughout the paper including a definition of anonymous identity-based encryption and testable identity-based encryption. In Sect. 3, we show the syntax and security of the new primitive anonymous identity-based encryption with identity recovery. In Sect. 4, we present the construction of AIBEIR and prove its security. We conclude the paper in Sect. 5.

\section{Preliminaries and Definitions}
We denote $s \xleftarrow{\$}\mathcal{S}$ as the operation of assigning to $s$ an element selected uniformly at random from set $\mathcal{S}$. The notation $x\leftarrow \textsf{A($\cdot$)}$ denotes the operation of running an algorithm \textsf{A} with some given input and assigning the output to $x$. A function \textsf{negl}: $\mathbb{N}\rightarrow \mathbb{R}$ is \emph{negligible} if for every positive polynomial \textsf{poly} and sufficiently large $\lambda$, it holds that \textsf{negl($\lambda$)}$<$$1/\textsf{poly($\lambda$)}$. We use $0^{\ell}$to denote the zero vector whose length is $\ell$. If $a$ is a vector, $|a|$ denotes its length.

\subsection{Bilinear Groups}
Let $\mathbb{G}_1$,$\mathbb{G}_2$ and $\mathbb{G}_T$ be multiplicative cyclic groups of prime order $p$. Let $g_1$,$g_2$ be generators of groups $\mathbb{G}_1$ and $\mathbb{G}_2$, respectively, and $e:\mathbb{G}_1\times \mathbb{G}_2 \rightarrow \mathbb{G}_T$ be a bilinear map that holds the following features:
\begin{itemize}
  \item Bilinearity: $e(u^a,v^b)=e(u,v)^{ab}$ for all $u\in \mathbb{G}_1$,$v\in \mathbb{G}_2$ and $a$,$b\in \mathbb{Z}_p$.
  \item Non-degeneracy: $e(g_1,g_2)\neq 1_{\mathbb{G}_T}$
  \item Computability: there exists an efficient algorithm to compute $e(u,v)$ for any input pair $u\in \mathbb{G}_1$,$v\in \mathbb{G}_2$.
\end{itemize}
We assume a symmetric bilinear map such that $\mathbb{G}_1=\mathbb{G}_2=\mathbb{G}$ and $g_1=g_2=g$.

\subsection{Identity-Based Encryption}

   Let $\lambda$ be a security parameter. An identity-based encryption is a tuple of algorithms $\Pi_{IBE}$ = \textsf{(IBE.Setup,IBE.Extract,IBE.Encrypt,IBE.Decrypt)} with the following properties:

   \begin{itemize}
     \item[$\bullet$]\textsf{Setup($1^{\lambda}$)}: This is a polynomial time algorithm which takes as input $1^{\lambda}$ and outputs the system parameter $mpk$ and a master secret key $msk$.

     \item[$\bullet$]\textsf{Extract($id,msk$):} This is a polynomial time algorithm which takes as input user's identity $id$ and master secret key $msk$, and outputs the user's corresponding private key $sk_{id}$.

     \item[$\bullet$]\textsf{Encrypt($m,id,mpk$):} This is a polynomial time algorithm which takes as input a message $m$ in the message space, system parameter $mpk$, the receiver's identity $id$ and outputs a ciphertext $c$ in the ciphertext space.

     \item[$\bullet$]\textsf{Decrypt($mpk,c,sk_{id}$):} This is a polynomial time algorithm which takes as input system parameter $mpk$, ciphertext $c$, user's private key $sk_{id}$, outputs the message $m$ in the message space.
\end{itemize}
    \textbf{Correctness.} We require correctness of decryption: that is, for all $\lambda$, all identity $id$ in the identity space, all $m$ in the specified message space, $Pr[\textsf{Decrypt($mpk$,$sk_{id}$,} \\\textsf{Encrypt($m,id,mpk$))}=m]=1-\textsf{negl($\lambda$)}$ holds, where the probability is taken over the randomness of the algorithms.\\
    \textbf{Anonymity and Semantic security.} When the ciphertext can not reveal information of the message, we say that the cryptosystem is chosen-plaintext secure. We say that the cryptosystem is anonymous if the ciphertext can not reveal information of the identity of the receiver. We combine these two notions.

 \begin{defn}
  An IBE scheme is anonymous against chosen-identity and chosen-plaintext attacks if there does not exist any polynomial adversary $\mathcal{A}$ who has non-negligible advantage in the following game:
 \end{defn}

  \textbf{Setup:} The challenger takes as input a security parameter $\lambda$ (in unary) and runs the \textsf{Setup} algorithm of the IBE. It provides $\mathcal{A}$ with the system parameters $mpk$ while keeping the master secret key $msk$ to itself.

  \vspace{1 ex}
  \textbf{Phase 1:} The adversary $\mathcal{A}$ can make any polynomial key-extraction queries defined as follows:
 key-extraction query ($id$): The adversary $\mathcal{A}$ can choose an identity $id$ and sends it to the challenger. The challenger generates a secret key $sk_{id}$ of $id$ and returns it to $\mathcal{A}$.

  \vspace{1 ex}
\textbf{Challenge:} When $\mathcal{A}$ decides that Phase 1 is complete, it chooses two equal-length plaintexts $m_0,m_1$ and two identities $id_0,id_1$ under the constraint that they have not been asked for the private keys. The challenger chooses uniformly at random two bits $b\in\{0,1\},\gamma\in\{0,1\}$ and sends a ciphertext $c^{*}$ of $m_b$ as the challenge ciphertext under $id_\gamma$ to $\mathcal{A}$.
\vspace{1 ex}

\textbf{Phase 2:} The adversary $\mathcal{A}$ can also make queries just like Phase 1 except that it cannot make a key-extraction query of either $id_0$ or $id_1$.

\vspace{1 ex}
\textbf{Guess:} $\mathcal{A}$ outputs a guess ($b^{\prime},\gamma^{\prime}$) of ($b,\gamma$).\\

We define the advantage of the adversary $\mathcal{A}$ as $Adv_{\mathcal{A}}$ = $|Pr[b = b^{\prime}\wedge\gamma = \gamma^{\prime}]-\frac{1}{4}|$.\\

\subsection{Testable Identity-Based Encryption}
\begin{defn}
   An Identity-Based Encryption is testable if there exists an algorithm \emph{\textsf{Test($\cdot,\cdot$)}} which takes as input $c$ and an identity $id$  and returns 1 if $c$ is a part of a valid cipertext under $id$ and 0 otherwise.
\end{defn}

\begin{rem}In our construction, we need the testable IBE to satisfy an additional property that the ciphertext $c$ can be partitioned into two parts $c_0$ and $c_1$ where $c_0$ contains information of the identity but no information of the message while $c_1$ contains information of the message but no information of the identity. And the test algorithm takes $c_0$ other than $c$ as input. Our construction works if the testable IBE without this property. But if it is satisfied, our construction is more efficient because encrypting the part containing information of plaintext is sufficient. Moreover, to our knowledge, all of the existing testable IBEs satisfy it. \end{rem}
Some realizations of IBE from bilinear maps such as \cite{DBLP:conf/eurocrypt/BonehB04a,Waters05} satisfy the definition of testable IBE. We will prove that the scheme in \cite{Waters05} is a testable IBE.\\

Let  $\mathbb{G}$ be a group of prime order, $p$, for which there exists an efficiently computable bilinear map into $\mathbb{G}_1$. Additionally, let  $e :\mathbb{G} \times \mathbb{G} \rightarrow \mathbb{G}_1$ denote the bilinear map and $g$ be the corresponding generator. The size of the group is determined by the security parameter. Identities will be represented as bit strings of length $n$, a separate parameter unrelated to $p$. The construction follows.

\textsf{Setup.}The system parameters are generated as follows. We choose a random generator, $g\in \mathbb{G}$ and $g_2$ randomly in $\mathbb{G}$. We choose a secret $\alpha$ $ \in$ $ \mathbb{Z}_p$ and set $g_1 = g^{\alpha}$. Further, choose a random value $u^{\prime}\in  \mathbb{G}$ and a random $n-$length vector $U = (u_i)$, whose elements are chosen at random from $\mathbb{G}$. The published public parameters are $g,g_1, g_2,u^{\prime},$ and $U$. The master secret key is $g^{\alpha}_2 $.\\

$\textsf{Key Generation.}$ Let $v$ be a $n$-bit string representing an identity, $v_i$ denote the $ith$ bit of $v$, and $\mathcal{V} \subseteq \{1, . . . , n\}$ be the set of all $i$ for which $v_i$ = 1. (That is $\mathcal{V}$ is the set of indices for which the bit string $v$ is set to 1.) A private key for identity $v$ is generated as follows. First, a random $r \in \mathbb{Z}_p$ is chosen. Then the private key is constructed as:

\begin{center}$d_v = (g^{\alpha}_2(u^{\prime}\prod \limits_{i \in \mathcal{V}}u_i)^{r},g^{r})$ \end{center}

\textsf{Encryption.} A message $M \in \mathbb{G}_1$ is encrypted for an identity $v$ as follows. A value $t \in \mathbb{Z}_p$ is chosen at random. The ciphertext is then constructed as:
\begin{center}$C= (e(g_1,g_2)^{t}M,g^{t},(u^{\prime}\prod \limits_{i \in \mathcal{V}}u_i)^{t})$ \end{center}

\textsf{Decryption.} Let $C = (C_1,C_2,C_3)$ be a valid encryption of $M$ under the identity $v$. Then $C$ can be decrypted by $d_v = (d_1,d_2)$ as:\\
$C_1\frac{e(d_2,C_3)}{e(d_1,C_2)}=(e(g_1,g_2)^{t}M)\frac{e(g^{r},(u^{\prime}\prod \limits_{i \in \mathcal{V}}u_i)^{t})}{e(g^{\alpha}_2(u^{\prime}\prod \limits_{i \in \mathcal{V}}u_i)^{r}),g^{t}})=(e(g_1,g_2)^{t}M)\frac{e(g,(u^{\prime}\prod \limits_{i \in \mathcal{V}}u_i)^{rt}))}{e(g_1,g_2)^{t}e((u^{\prime}\prod \limits_{i \in \mathcal{V}}u_i)^{rt},g)} \\= M$\\
We can also define a \textsf{Test} algorithm as follows:\\

\textsf{Test.}Let $C = (C_1,C_2,C_3)$ be a valid encryption under the identity $v$. Let $v^{\prime}$ be a $n$ bit string representing an identity, $v^{\prime}_i$ denote the $ith$ bit of $v^{\prime}$, and $\mathcal{V}^{\prime} \subseteq \{1, . . . , n\}$ be the set of all $i$ for which $v^{\prime}_i$ = 1. Output 1 if  $e(g,C_3)=e(C_2,(u^{\prime}\prod \limits_{i \in \mathcal{V}^{\prime}}u_i))$ and $\bot$ otherwise. In fact, ($C_1$,$C_2$) contain the information of the message and no information of the identity. $C_3$ contains information of the identity but no information of the message. So it is a testable IBE.\\

\section{Anonymous Identity-Based Encryption with Identity Recovery}

   Compared to identity based encryption, there is an additional algorithm Recover that takes recovery secret key and a ciphertext as inputs and output the corresponding identity or $\bot$ if the ciphertext is not valid. Let $\lambda$ be a security parameter. An anonymous identity-based encryption with recovery is a tuple of algorithms \textsf{$\Pi_{AIBEIR}$ = (AIBEIR.Setup,AIBEIR.Extract,AIBEIR.Encrypt,AIBEIR.Decrypt,\\AIBEIR.Recover)} with the following properties:

   \begin{itemize}
     \item[$\bullet$]\textsf{Setup($1^{\lambda}$:)}  This is a polynomial time algorithm which
takes as input $1^{\lambda}$ and outputs the system parameter $mpk$, a master secret key $msk$ and secret key of the identity recovery manager $sk_{IRM}$. Then PKG sends $sk_{IRM}$ to the identity recovery manager in a secret channel. It is operated by PKG.\\

     \item[$\bullet$]\textsf{Extract($id,msk$):} This is a polynomial time algorithm which takes as input a user' identity $id$ and $msk$, outputs the user' corresponding private key $sk_{id}$.\\

     \item[$\bullet$]\textsf{Encrypt($m,mpk,id$):} This is a polynomial time algorithm which takes as input a message $m$ in a specified message space, system parameter $mpk$, the receiver' identity $id$ and outputs a ciphertext $c$ in the ciphertext space. It is operated by the sender.\\

     \item[$\bullet$]\textsf{Decrypt($mpk,c,sk_{id}$):} This is a polynomial time algorithm which takes as input system parameter $mpk$, ciphertext $c$, user' private key $sk_{id}$, outputs the message $m$ in the message space. It is operated by the receiver.

         \vspace{1 ex}
     \item[$\bullet$]\textsf{Recover($c,sk_{IRM}$)}: The identity recovery manager outputs an identity $id$ if $c$ is a valid cipertext under $id$ and $\bot$ otherwise. It is operated by the identity recovery manager.

         \end{itemize}
$\textbf{Correctness.}$ We say that $\Pi_{AIBEIR}$ is correct if it satisfies the following two properties:
\begin{itemize}

\item[$\bullet$]$\textbf{Decryption correctness:}$ For any $id$ in identity space and $m$ in a specified message space, $Pr[\textsf{AIBEIR.Decrypt($sk_{id}$,AIBEIR.Encrypt($m$,$id$,$mpk$))}=m]=1-\textsf{negl($\lambda$)}$.

\item[$\bullet$]$\textbf{Recovery correctness:}$ For any valid ciphertext $c = \textsf{AIBEIR.Encrypt($m,id,$}\\\textsf{$mpk$)}$, $Pr[\textsf{Recover($sk_{IRM}$,$c$)}=id]=1-\textsf{negl($\lambda$)}$.

\end{itemize}
$\textbf{Anonymity.}$ The anonymity of AIBEIR is the same as that of anonymous IBE. Note that the recovery manager can not be the adversary.\\
$\textbf{Stronger semantic security.}$ In the semantic security model of IBE, adversary has no information about the master secret key $msk$. But in the definition of our AIBEIR scheme, the identity recovery manager holds $sk_{IRM}$ which makes it more powerful. So if the identity recovery manager is the adversary, the semantic security model of IBE is not feasible. We define a stronger semantic security as follows:

\begin{defn}
  An AIBEIR scheme is strongly semantic secure against chosen-identity and chosen-plaintext attacks if there does not exist any polynomial adversary $\mathcal{A}$ who have non-negligible advantage in the game below:
 \end{defn}

  \textbf{Setup:} The challenger takes as input a security parameter $\lambda$ (in unary) and runs the \textsf{Setup} algorithm of the AIBEIR. It provides $\mathcal{A}$ with the system public parameters $mpk$ and identity recovery secret key $sk_{IRM}$ while keeping the master secret key $msk$ to itself.\\

  \textbf{Phase 1:} The adversary $\mathcal{A}$ can make any polynomial key-extraction queries defined as follows:
 key-extraction query ($id$): $\mathcal{A}$ can choose an identity $id$ and send it to the challenger. The challenger generates secret key $sk_{id}$ and returns it to $\mathcal{A}$.\\

\textbf{Challenge:} When $\mathcal{A}$ decides that Phase 1 is complete, it chooses two equal-length plaintexts $m_0,m_1$ and an identity $id^{*}$ under the constraint that it has not asked for the private key and sends them to the challenger. The challenger chooses uniformly at random a bit $b\in\{0,1\}$ and sends a ciphertext $c^{*}$ = \textsf{Encrypt($m_{b},id^{*},mpk$)} as the challenge ciphertext to $\mathcal{A}$.\\

\textbf{Phase 2:} $\mathcal{A}$ can also make queries just like Phase 1 except that it cannot make a key-extraction query of $id^{*}$.\\

\textbf{Guess:} $\mathcal{A}$ outputs a guess $b^{\prime}$ of $b$.\\

We define the advantage of adversary $\mathcal{A}$ as $Adv_{\mathcal{A}}$ = $|Pr[b = b^{\prime}]-\frac{1}{2}|$\\

\noindent\textbf{Recovery.} An AIBEIR scheme is recoverable if \textsf{Recover} algorithm can always extract the right identity from a valid ciphertext and output $\bot$ when the input is an invalid ciphertext.
\begin{defn}
  An AIBEIR scheme is recoverable if there does not exist any PPT adversary $\mathcal{A}$ who wins the following game with non-negligible probability.
 \end{defn}

\textbf{Setup:} The challenger takes as input a security parameter $\lambda$ (in unary) and runs the \textsf{Setup} algorithm of the AIBEIR. It provides $\mathcal{A}$ with the system parameters $mpk$ while keeping the master secret key $msk$ and $sk_{IRM}$ to itself.

\vspace{1 ex}
 \textbf{Monitor Phase:} The adversary $\mathcal{A}$ can query recover oracle and key-extraction oracle.

 \vspace{1 ex}
 \textbf{Challenge:} When $\mathcal{A}$ decides that Monitor Phase is complete, the adversary sends $c^{*}$ to the challenger. The challenger sends the output of \textsf{Recover} algorithm to $\mathcal{A}$.

 \vspace{1 ex}
 \textbf{Output:} $\mathcal{A}$ wins the game if the output of \textsf{Recover($c^{*},sk_{IRM}$)} is $\bot$ or $id$ while $c^{*}$ is a valid ciphertext under $id^{\prime}$ where $id\neq id^{\prime}$ or the output of \textsf{Recover($c^{*},sk_{IRM}$)} is $id$ while $c^{*}$ is not a valid ciphertext. Here we require $id$ has not been asked as a key-extraction query for the need to prove the security.

\section{A Construction from Anonymous IBE and Testable IBE}
In this section, we present our construction of AIBEIR from anonymous IBE and testable IBE. Let $\Pi_{1}$ = \textsf{(A-IBE.Setup,A-IBE.Enc,A-IBE.Dec,A-IBE.Extract)} be an anonymous IBE scheme, $\Pi_{2}$ = \textsf{(T-IBE.Setup,T-IBE.Enc,T-IBE.Dec,T-IBE.Extract,\\T-IBE.Test)} be a testable IBE scheme. Let $id_{\epsilon}$ denote the identity of the identity recovery manager in scheme $\Pi_{2}$. Then, we can construct an AIBEIR scheme $\Pi$ as follows:

\vspace{1 ex}
\subsection{The Construction} We describe our AIBEIR scheme \textsf{(AIBEIR.Setup, AIBEIR.Extract, AIBEIR.Encrypt,\\AIBEIR.Decrypt,AIBEIR.Recover)} as follows:
\begin{itemize}
     \item[$\bullet$]\textsf{Setup($1^{\lambda}$)}: Run the \textsf{Setup} algorithms of \textsf{A-IBE} and \textsf{T-IBE} and obtain $(MPK_A,\\MSK_A)\leftarrow $ \textsf{A-IBE.Setup($1^{\lambda}$)} , $ (MPK_T,MSK_T) \leftarrow $\textsf{T-IBE.Setup($1^{\lambda}$)} , respectively. Compute $SK_{T,id_{\epsilon}}$ = \textsf{T-IBE.Extract($MSK_T,id_{\epsilon}$)}. $(mpk,msk) = ((MPK_A,MPK_T),(MSK_A,MSK_T)) $ ,$sk_{IRM}$ = ($MSK_A,SK_{T,id_{\epsilon}}$).

     \item[$\bullet$]\textsf{Extract($id,msk$)}: Run the \textsf{Extract} algorithms of \textsf{A-IBE} and \textsf{T-IBE} and obtain $SK_{A,id}$ = \textsf{A-IBE.Extract($id,MSK_A$)} and $SK_{T,id}$ = \textsf{T-IBE.Extract($id,MSK_T$)}, respectively. Output $sk_{id} = (SK_{A,id},SK_{T,id})$.

     \item[$\bullet$]\textsf{Encrypt($m,id,mpk$):} Run the \textsf{Encrypt} algorithms of \textsf{A-IBE} and \textsf{T-IBE} and obtain $(c_0,c_1)$ = \textsf{T-IBE.Enc($m,id,MPK_T$)}, $c_2$ = \textsf{A-IBE.Enc($c_0,id,$}\\\textsf{$MPK_A$)} and $c_3$ = \textsf{T-IBE.Enc($id,id_{\epsilon},MPK_T$)}. Output $c$ = ($c_1,c_2,c_3$).

     \item[$\bullet$]\textsf{Decrypt($mpk,c,sk_{id}$):} Parse $c$ as $c_1,c_2$ and $c_3$. Then compute $c_0$ = \textsf{A-IBE.Dec(}\\\textsf{$c_2,SK_{A,id}$)}, $m$ = \textsf{T-IBE.Dec($c_0||c_1,SK_{T,id}$)}.

     \item[$\bullet$]\textsf{Recover($c,sk_{IRM}$)}: Parse $c$ as $c_1,c_2$ and $c_3$. Parse $sk_{IRM}$ as $MSK_A$ and $SK_{T,id_{\epsilon}}$. Then compute $id$ = \textsf{T-IBE.Dec($c_3,SK_{T,id_{\epsilon}}$)} and $SK_{A,id}$ = \textsf{A-IBE.Ext}\\\textsf{ract($id,MSK_A$)}. Take as input $SK_{A,id}$ and $c_2$ , obtain the cipertext $c_0$ by running the \textsf{Decrypt} algorithm of \textsf{A-IBE.Dec($SK_{A,id}$, $c_2$)}. Finally, output $id$ if \textsf{T-IBE.Test($id,c_0$)} = 1, and $\bot$ otherwise.
         \end{itemize}
\begin{rem}
         Here the message space of $\Pi_1$ includes the ciphertext space of $\Pi_2$. We set the intersection of identity space of $\Pi_1$ and $\Pi_2$ as the identity space of $\Pi$.
         \end{rem}
\subsection{Correctness}
\begin{thm}
If $\Pi_{1}$ is a correct anonymous IBE scheme and $\Pi_{2}$ is a correct testable IBE scheme then $\Pi$ is a correct AIBEIR scheme.
\end{thm}
\begin{itemize}
\item[$\bullet$]\textbf{Decryption correctness:} The decryption correctness is guaranteed by the decryption correctness of $\Pi_1$ and $\Pi_2$.
\item[$\bullet$]\textbf{Recovery correctness:} The recovery correctness is guaranteed by the decryption correctness of $\Pi_1$ , $\Pi_2$ and test correctness of $\Pi_2$.
\end{itemize}
\subsection{Anonymity}
\begin{thm}
If $\Pi_{1}$ is an IBE scheme which is anonymous against adaptively chosen-identity and chosen-plaintext attacks and $\Pi_{2}$ is a testable IBE scheme which is fully secure against chosen-identity and chosen-plaintext attacks, then $\Pi$ is an AIBEIR scheme which is anonymous against adaptively chosen-identity and chosen-plaintext attacks.\footnote{Here the adversary can not be the identity recovery manager and has PPT power. If AIBE and TIBE are both selective secure, our AIBEIR scheme is also selective secure.}
\end{thm}

\textit{Proof}. We prove the above theorem by hybrid arguments.

\vspace{1 ex}
$\mathcal{H}_0$: This hybrid is the real experiment in the Definition 1. The logic of the challenger is shown as follows:\\
initialization:
\begin{itemize}
\item[] $(MPK_A,MSK_A)\leftarrow $ \textsf{A-IBE.Setup($1^{\lambda}$)} , $ (MPK_T,MSK_T) \leftarrow $\textsf{T-IBE.Setup($1^{\lambda}$)}
\item[]$(mpk,msk) =  ((MPK_A,MPK_T),(MSK_A,MSK_T)) $
\item[] $SK_{T,id_{\epsilon}}$ = \textsf{T-IBE.Extract($MSK_T,id_{\epsilon}$)}, $sk_{IRM}$ = ($MSK_A,SK_{T,id_{\epsilon}}$)
\item[] send $mpk$ to $\mathcal{A}$
\end{itemize}
upon receiving a secret key query($id$):
\begin{itemize}
\item[] $SK_{A,id}$ = \textsf{A-IBE.Extract($id,MSK_A$)} and $SK_{T,id}$ = \textsf{T-IBE.Extract($id,MSK_T$)}
\item[] send $sk_{id} = (SK_{A,id},SK_{T,id})$ to $\mathcal{A}$
\end{itemize}
upon receiving the challenge query ($m_0, m_1,id_0,id_1$):
\begin{itemize}
\item[]    $b\xleftarrow{\$}\{0,1\}$ ,$\gamma \xleftarrow{\$}\{0,1\}$,
\item[(1)] $(c_0,c_1)$ = \textsf{T-IBE.Enc($m_b,id_{\gamma},MPK_T$)}
\item[(2)] $c_2$ = \textsf{A-IBE.Enc($c_0,id_{\gamma},MPK_A$)}
\item[(3)] $c_3$ = \textsf{T-IBE.Enc($id_{\gamma},id_{\epsilon},MPK_T$)}
\item[] send $c$ = ($c_1,c_2,c_3$) to $\mathcal{A}$
\end{itemize}
$\mathcal{H}_1$: In this hybrid, it is identical to $\mathcal{H}_0$ except that we just change how the challenge ciphertext is generated. We replace the lines marked (1) in $\mathcal{H}_0$ as follows:

\begin{itemize}
\item[] $c_0,c_1$ = \textsf{T-IBE.Enc($0^{|m_b|},id_{\gamma},MPK_T$)}.
\end{itemize}
$\mathcal{H}_2$: Compared to $\mathcal{H}_1$, we replace the lines marked (2) in $\mathcal{H}_0$ as follows:
\begin{itemize}
\item[] $c_2$ =\textsf{ A-IBE.Enc($0^{|c_0|},id_{\gamma},MPK_A$)}.

\end{itemize}
$\mathcal{H}_3$: Same as $\mathcal{H}_2$, except we replace the lines marked (2) in $\mathcal{H}_0$ as follows:
  \begin{itemize}
\item[]We just randomly choose $id$ from identity space except $id_0$ and $id_1$. We then set $c_2$ = \textsf{A-IBE.Enc($0^{|c_0|},id,MPK_A$)}.
\end{itemize}
\vspace{1 ex}
 $\mathcal{H}_4$: Identical to $\mathcal{H}_3$, except we replace the lines marked (3) in $\mathcal{H}_0$ as follows:
 \begin{itemize}
\item[] We just set $c_3$ as \textsf{T-IBE.Enc($0^{|id_{\gamma}|},id_{\epsilon},MPK_T$)}.
\end{itemize}
It is easy to know that the challenge ciphertext in $\mathcal{H}_4$ contains no information about $b$ and $\gamma$ (except their length). So the advantage of $\mathcal{A}$ in $\mathcal{H}_4$ is 0. We prove the above theorem by showing that $\mathcal{H}_0\approx\mathcal{H}_1\approx\mathcal{H}_2\approx\mathcal{H}_3\approx\mathcal{H}_4 $ through the following lemmas.
 \vspace{2 ex}
 \begin{lem}
 Any $PPT$ adversary cannot distinguish $\mathcal{H}_0$ and $\mathcal{H}_1$, if scheme $\Pi_2$ is fully secure against adaptively chosen-identity and chosen-plaintext attacks.
 \end{lem}
 \emph{Proof.} We can construct a simulator $\mathcal{B}$ to break the full security against chosen-identity and chosen-plaintext attacks of scheme $\Pi_2$, if there is an adversary $\mathcal{A}$ who can distinguish $\mathcal{H}_0$ and $\mathcal{H}_1$.

\textbf{Setup:} The challenger takes as input a security parameter $\lambda$ (in unary) and runs the \textsf{Setup} algorithm of $\Pi_{2}$. It provides $\mathcal{B}$ with the system parameters $MPK_T$ while keeping the master secret key $MSK_T$ to itself. $\mathcal{B}$ computes $(MPK_A,MSK_A)\leftarrow $ \textsf{A-IBE.Setup($1^{\lambda}$)}, and sends $MPK = (MPK_A,MPK_T) $ to $\mathcal{A}$.

 \vspace{1 ex}
\textbf{Phase 1:} When the adversary $\mathcal{A}$ makes key-extraction query and sends an identity $id$ to $\mathcal{B}$, $\mathcal{B}$ just forwards it as the key-extraction query to the challenger. The challenger sends $SK_{T,id}$ to $\mathcal{B}$. $\mathcal{B}$ computes $SK_{A,id}$ = \textsf{A-IBE.Extract($id,MSK_A$)} and sends $sk_{id} = (SK_{A,id},SK_{T,id})$ to $\mathcal{A}$.\\

\textbf{Challenge:} $\mathcal{A}$ chooses $id_{0}$ and $id_{1}$  under the constraint that they have not been asked for the private keys and two equal-length messages $m_0,m_1$ and sends them to $\mathcal{B}$. $\mathcal{B}$  just chooses randomly two bits $b$ and $\gamma$ and sends ($m_b,\textbf{0},id_{\gamma}$) to the challenger. The challenger chooses uniformly at random a bit $b^{\prime}$ and sends $c_0,c_1$ = \textsf{T-IBE.Enc($m,id_{\gamma},MPK_T$)} to $\mathcal{B}$. If $b^{\prime} = 0$, $m = m_b$. If $b^{\prime} = 1$,$m = \textbf{0}$. $\mathcal{B}$ obtains $c_2,c_3$ by running \textsf{A-IBE.Enc($c_0,id_{\gamma},MPK_A$)} and \textsf{T-IBE.Enc($id_{\gamma},id_{\epsilon},MPK_T$)} respectively.  $\mathcal{B}$ just sends $c^{*} =(c_1,c_2,c_3 )$ to $\mathcal{A}$. \\

\textbf{Phase 2:} $\mathcal{A}$ makes key-extraction queries except $id_{0},id_{1}$. $\mathcal{B}$ answers queries just like Phase 1.\\

\textbf{Guess} $\mathcal{A}$ sends a bit $\bar{b}$ as a guess of $\mathcal{H}_{\bar{b}}$ to $\mathcal{B}$. $\mathcal{B}$ just forwards it to the challenger.\\

The view of $\mathcal{A}$ is identical to $\mathcal{H}_0$ if $b^{\prime}= 0$ and to $\mathcal{H}_1$ if $b^{\prime}= 1$. Thus, by the semantic security of scheme $\Pi_2$, we can conclude that $\mathcal{H}_0 \approx \mathcal{H}_1$.
\begin{lem}
 Any $PPT$ adversary cannot distinguish $\mathcal{H}_1$ and $\mathcal{H}_2$, if scheme $\Pi_1$ is anonymous against adaptive-identity, chosen-plaintext attacks.
 \end{lem}
 \textit{Proof}. Given a PPT adversary $\mathcal{A}$ who can distinguish $\mathcal{H}_1$ and $\mathcal{H}_2$, we can construct a simulator $\mathcal{B}$ attacking the anonymous security of $\Pi_1$ against adaptive-identity, chosen-plaintext attacks.\\

\textbf{Setup:} The challenger takes as input a security parameter $\lambda$ (in unary) and runs the \textsf{Setup} algorithm of $\Pi_{1}$. It provides $\mathcal{B}$ with the system parameters $MPK_A$ while keeping the master secret key $MSK_A$ to itself. $\mathcal{B}$ computes $ (MPK_T,MSK_T) \leftarrow$ \textsf{T-IBE.Setup($1^{\lambda}$)}, and sends $MPK = (MPK_A,MPK_T) $ to $\mathcal{A}$.\\

\textbf{Phase 1:} When $\mathcal{A}$ makes key-extraction query and sends an identity $id$ to $\mathcal{B}$, $\mathcal{B}$ just forwards $id$ as the key-extraction query to the challenger. The challenger sends $SK_{A,id}$ to $\mathcal{B}$. $\mathcal{B}$ runs $SK_{T,id}$ = \textsf{T-IBE.Extract($id,MSK_T$)} and sends $sk_{id} = (SK_{A,id},SK_{T,id})$ to $\mathcal{A}$.\\

\textbf{Challenge:} $\mathcal{A}$ chooses two equal-length plaintexts $m_0,m_1$ and two identities $id_0,id_1$ under the constraint that they have not been asked for the private keys and sends them to $\mathcal{B}$. $\mathcal{B}$  chooses uniformly at random a bit $\gamma^{\prime}\in\{0,1\}$ and computes $c_0,c_1$ = \textsf{T-IBE.Enc($\textbf{0},id_{\gamma^{\prime}},MPK_T$)}, $c_3$ = \textsf{T-IBE.Enc($id_{\gamma^{\prime}},id_{\epsilon},MPK_T$)}. $\mathcal{B}$ sends ($c_0$,\textbf{0},$id_{\gamma^{\prime}}$,$id_{\gamma^{\prime}}$ )to the challenger. The challenger chooses uniformly at random a bit $\gamma$ and a bit $b$. If $b$ =0, the challenger sends $c_2$ = \textsf{A-IBE.Enc($c_0,id_{\gamma^{\prime}},MPK_A$)} to $\mathcal{B}$. If $b$ =1, the challenger sends $c_2$ = A-IBE.Enc($\textbf{0},id_{\gamma^{\prime}},MPK_A$) to $\mathcal{B}$. $\mathcal{B}$  sends ($c_1,c_2,c_3)$ to $\mathcal{A}$.\\

\textbf{Phase 2:} $\mathcal{B}$ answers queries just like Phase 1, but $id_0$ and $id_1$ cannot be queried.\\

\textbf{Guess:} $\mathcal{A}$ sends a bit $\bar{b}$ as a guess of $\mathcal{H}_{\bar{b}+1}$ to $\mathcal{B}$. $\mathcal{B}$ randomly choose a bit $\gamma$ and sends $\bar{b}$ and $\gamma$ to the challenger.\\

 If $b = 0$, the view of $\mathcal{A}$ is identical to $\mathcal{H}_1$. If  $b = 1$, the view of $\mathcal{A}$ is identical to $\mathcal{H}_2$. We can see that $\mathcal{H}_1 \approx \mathcal{H}_2$ by the anonymity of $\Pi_1$.

\begin{lem}
 Any $PPT$ adversary cannot distinguish $\mathcal{H}_2$ and $\mathcal{H}_3$, if scheme $\Pi_1$ is anonymous secure against adaptively chosen-identity, chosen-plaintext attacks.
 \end{lem}
 \textit{Proof}. Given a PPT adversary $\mathcal{A}$ who can distinguish $\mathcal{H}_2$ and $\mathcal{H}_3$, we can construct a simulator $\mathcal{B}$ attacking the anonymous security of $\Pi_1$ against adaptively chosen-identity, chosen-plaintext attacks.\\

\textbf{Setup:} The challenger takes as input a security parameter $\lambda$ (in unary) and runs the \textsf{Setup} algorithm of $\Pi_{1}$. It provides $\mathcal{B}$ with the system parameters $MPK_A$ while keeping the master secret key $MSK_A$ to itself. $\mathcal{B}$ computes $ (MPK_T,MSK_T) \leftarrow $\textsf{T-IBE.Setup($1^{\lambda}$)}, and sends $mpk = (MPK_A,MPK_T) $ to $\mathcal{A}$.\\

\textbf{Phase 1:} When the adversary $\mathcal{A}$ makes key-extraction query and sends an identity $id$ to $\mathcal{B}$, $\mathcal{B}$ just forwards $id$ as the key-extraction query to the challenger. The challenger sends $SK_{A,id}$ to $\mathcal{B}$. $\mathcal{B}$ obtains $SK_{T,id}$ = \textsf{T-IBE.Extract($id$,}\textsf{$MSK_T$)} and sends $sk_{id} = (SK_{A,id},SK_{T,id})$ to $\mathcal{A}$.\\

\textbf{Challenge:} $\mathcal{A}$ chooses two equal-length plaintexts $m_0,m_1$ and two identities $id_0,id_1$ under the constraint that they have not been asked for the private keys and sends them to $\mathcal{B}$. $\mathcal{B}$  chooses uniformly at random a bit $\gamma^{\prime}\in\{0,1\}$ and computes $c_0,c_1$ = \textsf{T-IBE.Enc($\textbf{0},id_{\gamma^{\prime}},MPK_T$)}, $c_3$ = \textsf{T-IBE.Enc($id_{\gamma^{\prime}},id_{\epsilon},MPK_T$)}. $\mathcal{B}$ randomly chooses an identity $id$ from identity space except $id_0$, $id_1$ and sends ($\textbf{0},\textbf{0}$,$id_{\gamma^{\prime}}$, $id$ )to the challenger. The challenger chooses uniformly at random a bit $\gamma$ and a bit $b$.If $\gamma$ =0 ,the challenger sends $c_2$ = \textsf{A-IBE.Enc($\textbf{0},id_{\gamma^{\prime}},MPK_A$) to $\mathcal{B}$}. If $\gamma$ =1 ,the challenger sends $c_2$ = \textsf{A-IBE.Enc($\textbf{0},id,MPK_A$)} to $\mathcal{B}$. $\mathcal{B}$  sends ($c_1,c_2,c_3)$ to $\mathcal{A}$.\\

\textbf{Phase 2:} $\mathcal{B}$ answers queries just like Phase 1, but $id_0$ and $id_1$ cannot be asked.\\

\textbf{Guess:} $\mathcal{A}$ sends a bit $\bar{\gamma}$ as a guess of $\mathcal{H}_{\bar{\gamma}+2}$  to $\mathcal{B}$. $\mathcal{B}$ randomly choose a bit $\bar{b}$ and sends $\bar{\gamma}$ and $\bar{b}$ to the challenger.

If $\gamma = 0$, the view of $\mathcal{A}$ is identical in $\mathcal{H}_2$. If  $\gamma = 1$, the view of $\mathcal{A}$ is identical in $\mathcal{H}_3$.The probability that $\mathcal{A}$ can distinguish $\mathcal{H}_2$ and $\mathcal{H}_3$ equals $|Pr[\bar{\gamma}= \gamma]-\frac{1}{2}|$ = $|2(\frac{1}{4}+\textsf{negl(n)})-\frac{1}{2}|$ = \textsf{negl(n)} because of the anonymity of $\Pi_1$. So the conclusion is that $\mathcal{H}_2 \approx \mathcal{H}_3$.

\begin{lem}
 Any $PPT$ adversary cannot distinguish $\mathcal{H}_3$ and $\mathcal{H}_4$, if scheme $\Pi_2$ is secure against chosen-identity and chosen-plaintext attacks.
 \end{lem}
 \textit{Proof}. Given a $PPT$ adversary $\mathcal{A}$ which can distinguish $\mathcal{H}_3$ and $\mathcal{H}_4$, we can construct a simulator $\mathcal{B}$ attacking the semantic security of $\Pi_2$ against chosen-identity and chosen-plaintext attacks.\\

\textbf{Setup:} The challenger takes as input a security parameter $\lambda$ (in unary) and runs the \textsf{Setup} algorithm of $\Pi_{2}$ and obtains ($MPK_T,MSK_T$). It sends $MPK_T$ to $\mathcal{B}$ and keeps $MSK_T$ to itself. $\mathcal{B}$ computes $(MPK_A,MSK_A)\leftarrow $ \textsf{A-IBE.Setup($1^{\lambda}$)} and sends $mpk = (MPK_A,MPK_T) $ to $\mathcal{A}$.\\

\textbf{Phase 1:} When the adversary $\mathcal{A}$ makes key-extraction query and sends an identity $id$ to $\mathcal{B}$, $\mathcal{B}$ just forwards it as the key-extraction query to the challenger. The challenger sends $MSK_{T,id}$ to $\mathcal{B}$. $\mathcal{B}$ computes $SK_{A,id}$ = \textsf{A-IBE.Extract($id,MSK_A$)} and sends $sk_{id} = (SK_{A,id},SK_{T,id})$ to $\mathcal{A}$.\\

\textbf{Challenge:} $\mathcal{A}$ chooses two equal-length plaintexts $m_0,m_1$ and two identities $id_0,id_1$ under the constraint that they have not been asked for the private keys and sends them to $\mathcal{B}$. $\mathcal{B}$  chooses uniformly at random a bit $\gamma\in\{0,1\}$ and computes $c_0,c_1$ = \textsf{T-IBE.Enc($\textbf{0},id_{\gamma},MPK_T$)}. $\mathcal{B}$ randomly chooses an identity $id$ from the identity space except $id_0$, $id_1$ and  computes $c_2$ = \textsf{A-IBE.Enc($\textbf{0},id,MPK_A$)}. $\mathcal{B}$  sends ($id_{\gamma},\textbf{0},id_{\epsilon}$)to the challenger.The challenger chooses uniformly at random a bit $b$ and sends $c_3$ to $\mathcal{B}$. $c_3$ = \textsf{T-IBE.Enc($id_{\gamma},id_{\epsilon},MPK_T$)}, if $b = 0$. $c_3$ = T-IBE.Enc($\textbf{0},id_{\epsilon},MPK_T$), if $b = 1$. $\mathcal{B}$ just sends $c^{*} =(c_1,c_2,c_3 )$ to $\mathcal{A}$. \\

\textbf{Phase 2:} $\mathcal{A}$ makes key-extraction queries except $id_{0},id_{1}$. $\mathcal{B}$ answers queries just like Phase 1.

The view of $\mathcal{A}$ is identical to $\mathcal{H}_3$ if $b= 0$, and $\mathcal{H}_4$ otherwise. The probability that the adversary can distinguish $\mathcal{H}_3$ and $\mathcal{H}_4$ equals the advantage of $\mathcal{B}$ breaking the semantic security of $\Pi_2$. So we can draw the conclusion that $\mathcal{H}_3 \approx \mathcal{H}_4$.

Having proved the above lemmas, we have completed the proof of Theorem 2.

\subsection{Stronger Semantic Security}

\begin{thm}
The AIBEIR scheme $\Pi$ is strongly semantic secure if $\Pi_2$ is semantic secure against chosen-identity and chosen-plaintext attack.
\end{thm}
\emph{Proof.} We can construct a simulator $\mathcal{B}$ breaking semantic security of $\Pi_2$ if there exists an adversary $\mathcal{A}$ breaking the stronger semantic security of $\Pi$.\\

\textbf{Setup:} The challenger takes as input a security parameter $\lambda$ (in unary) and runs the \textsf{Setup} algorithm of $\Pi_{2}$ and obtains ($MPK_T,MSK_T$). It sends $MPK_T$ to $\mathcal{B}$ and keeps $MSK_T$ to itself. $\mathcal{B}$ computes $(MPK_A,MSK_A) \leftarrow$ \textsf{A-IBE.Setup($1^{\lambda}$)}. $\mathcal{B}$ obtains $SK_{T,id_{\epsilon}}$ by making the secret key query of $id_{\epsilon}$ to the challenger and sends $mpk = (MPK_A,MPK_T) $ and $sk_{IRM}=(MSK_A,SK_{T,id_{\epsilon}})$ to $\mathcal{A}$.\\

\textbf{Phase 1:} When the adversary $\mathcal{A}$ makes key-extraction query and sends an identity $id$ to $\mathcal{B}$, $\mathcal{B}$ just forwards it as the key-extraction query to the challenger. The challenger sends $MSK_{T,id}$ to $\mathcal{B}$. $\mathcal{B}$ computes $SK_{A,id}$ = \textsf{A-IBE.Extract($id,MS$}\\\textsf{$K_A$)} and sends $sk_{id} = (SK_{A,id},SK_{T,id})$ to $\mathcal{A}$.\\

\textbf{Challenge:} $\mathcal{A}$ chooses two equal-length plaintexts $m_0,m_1$ and an identity $id^{*}$ under the constraint that it has not been asked for the private key and sends them to $\mathcal{B}$. $\mathcal{B}$  just forwards ($m_0, m_1, id^{*}$) to the challenger. The challenger randomly chooses a bit $b$ and sends ($c^{*}_0$, $c^{*}_1$) = \textsf{T-IBE.Enc($m_b,id^{*},MPK_T$)}. $\mathcal{B}$ computes $c^{*}_2$ = \textsf{A-IBE.Enc($c^{*}_0,$ $id^{*},MPK_A$)}, $c^{*}_3$ = \textsf{T-IBE.Enc($id^{*},id_{\epsilon},MPK_T$)} and sends $c^{*}$ = ($c^{*}_1$, $c^{*}_2$, $c^{*}_3$) to $\mathcal{A}$.\\

\textbf{Phase 2:} $\mathcal{A}$ makes key-extraction queries except $id^{*}$. $\mathcal{B}$ answers queries just like Phase 1.\\

\textbf{Guess:} $\mathcal{B}$ just forwards the output of $\mathcal{A}$ to the challenger.\\

If $\mathcal{A}$ wins, we can see $\mathcal{A}$ as a distinguish oracle. When $\mathcal{B}$ obtains the challenge ciphertext from challenger, $\mathcal{B}$ just encrypts it by the\textsf{ Encrypt} algorithm of \textsf{A-IBE} and sends it to $\mathcal{A}$. We can see that the probability that $\mathcal{A}$ breaks the stronger semantic security equals the probability that $\mathcal{B}$ breaks the semantic security of $\Pi_2$.
\subsection{Recovery}
\begin{thm}
  If the testable IBE scheme $\Pi_2$ is secure against adaptive-identity and chosen ciphertext attack, then the AIBEIR scheme $\Pi$ satisfies recovery.
\end{thm}
\emph{Proof.} If the adversary wins in the recovery experiment, there are two cases: (1) the adversary outputs a valid AIBEIR ciphertext but the challenger output $\bot$ or a wrong identity. This will not happen, which is guaranteed by the correctness of \textsf{Recover} algorithm. (2) the adversary outputs an invalid AIBEIR ciphertext but the challenger does not output $\bot$. We just consider the case where ($c_1,c_2,c_3$) is a valid ciphertxt \footnote{Although $c_1,c_2,c_3$ are all valid ciphertext, it maybe not a valid AIBEIR ciphertext. Note that an AIBEIR ciphertext ($c_1,c_2,c_3$) is valid where $c_1,c_2,c_3$ are valid ciphertexts under the same identity. }. In fact, if ($c_1,c_2$) is not a valid ciphertext, the receiver cannot decrypt correctly using its secret key. And if $c_3$ is not a valid T-IBE ciphertext under $id_{\epsilon}$, challenger will output $\bot$.

If ($c_1,c_2$) is a valid ciphertext under $id$ and $c_3$ is a testable IBE ciphertext of a different identity $\widehat{id}$ under $id_{\epsilon}$, we can show that the identity recovery manager will return $\bot$ with overwhelming probability. In fact, if there exists a PPT adversary $\mathcal{A}$ who can fool the identity recovery manager in the recovery game, we can construct a simulator $\mathcal{S}$ attacking  $\Pi_2$ in adaptive-identity, chosen-plaintext attack.

\textbf{Setup:} The challenger takes as input a security parameter $\lambda$ (in unary) and runs the \textsf{Setup} algorithm of $\Pi_{2}$. It provides $\mathcal{B}$ with the system parameters $MPK_T$ while keeping the master secret key $MSK_T$ to itself. $\mathcal{B}$ computes $(MPK_A,MSK_A)\leftarrow $ \textsf{A-IBE.Setup($1^{\lambda}$)} and sends $mpk = (MPK_A,MPK_T) $ to $\mathcal{A}$.

\vspace{1 ex}
\textbf{Phase 1:} When the adversary $\mathcal{A}$ makes the key-extraction queries, $\mathcal{B}$ just forwards the identity queried by $\mathcal{A}$ to the challenger and obtains $SK_{T,id}$ from the challenger. $\mathcal{B}$ obtains $SK_{A,id}$ = \textsf{A-IBE.Extract($id,MSK_A$)} and sends $sk_{id} = (SK_{A,id},SK_{T,id})$ to $\mathcal{A}$. When $\mathcal{A}$ makes recover query, $\mathcal{B}$ gets $SK_{T,id_{\epsilon}}$ by making secret key query of $id_{\epsilon}$ to the challenger and obtains $id$ by decrypting $c_3$ using $SK_{T,id_{\epsilon}}$. $\mathcal{B}$ computes  $SK_{A,id}$ = \textsf{A-IBE.Extract($id,MSK_A$)} and then obtains  $c_0$ by running \textsf{Dec} algorithm of \textsf{A-IBE}. $\mathcal{B}$ computes $h$ = \textsf{T-IBE.Test($c_0,id$)}, and sends $id$ to $\mathcal{A}$ if $h$ = 1, and $\bot$ otherwise. We say $\mathcal{A}$ wins if it outputs a valid ``double-encrypt'' IBE ciphertext ($c_1,c_2$) under $id_1$ and a valid testable IBE ciphertext $c_3$ of $id_2$ which pass the recover algorithm\footnote{This means we can obtain a T-IBE ciphertext under $id_2$ by decrypting the ``double-encrypt'' ciphertext under $id_1$ using $SK_{A,id_2}$}($\mathcal{A}$ can output the randomness used in the encrypt algorithm to show it ). Here we constrain that $id_1$ has not been queried the private key before. $\mathcal{B}$ obtains $SK_{T,id_2}$ by making the secret key query of $id_2$.

\vspace{1 ex}
\textbf{Challenge:} $\mathcal{B}$ randomly chooses two equal-length message $m_0,m_1$ and sends $m_0,m_1$ and $id_1$ to challenger. Challenger randomly chooses a bit $b\in\{0,1\}$ and obtains $(c_0,c_1)$ = \textsf{T-IBE.Enc($m_b,id_1,MPK_T$)}.

\vspace{1 ex}
\textbf{Phase 2:} $\mathcal{B}$ makes some queries to key-extraction oracle. In fact, $\mathcal{B}$ does not need to query now.

\vspace{1 ex}
\textbf{Guess:} $\mathcal{B}$ computes $c_2$ = \textsf{A-IBE.Enc($c_0,id_1,MPK_A$)} and obtains $c^{\prime}_0$ which is a part of ciphertext under $id_2$ by decrypting $c_2$ using $SK_{A,id_2}$. Then $\mathcal{B}$ obtains $m$ by decrypting $c^{\prime}_0$, $c_1$ using $SK_{T,id_2}$. $\mathcal{B}$ outputs 0 if $m$ = $m_0$ and 1 otherwise.

\section{Conclusion}
We define a new primitive called AIBEIR and construct it using double encryption with an anonymous IBE and a testable IBE. AIBEIR is anonymous for all users except the identity recovery manager who can recover the identity from the ciphertext. But the identity recovery manager can not obtain information about plaintext from ciphertext even holding an identity recover secret key. To our knowledge, \cite{DBLP:conf/eurocrypt/BonehB04a,Waters05} and their variations satisfy our testable IBE definition. We leave as an open problem the question of constructing testable IBE from other standard assumptions, such as lattice. Another interesting area of research is to construct more practical AIBEIR schemes.

\section*{Acknowledgements.} We would like to thank the anonymous reviewers of ACISP 2018 for their advice. Xuecheng Ma and Dongdai Lin are supported by the National Natural Science Foundation of China under Grant No. 61379139.
\bibliographystyle{alpha}
\bibliography{AIBEIR}

\newcommand{\etalchar}[1]{$^{#1}$}
\begin{thebibliography}{CHKP10}

\bibitem[ABB10]{ABB10a}
Shweta Agrawal, Dan Boneh, and Xavier Boyen.
\newblock Efficient lattice {(H)IBE} in the standard model.
\newblock In {\em Advances in Cryptology - {EUROCRYPT} 2010, 29th Annual
  International Conference on the Theory and Applications of Cryptographic
  Techniques, French Riviera, May 30 - June 3, 2010. Proceedings}, pages
  553--572, 2010.

\bibitem[AG09]{DBLP:conf/ctrsa/AtenieseG09}
Giuseppe Ateniese and Paolo Gasti.
\newblock Universally anonymous {IBE} based on the quadratic residuosity
  assumption.
\newblock In {\em Topics in Cryptology - {CT-RSA} 2009, The Cryptographers'
  Track at the {RSA} Conference 2009, San Francisco, CA, USA, April 20-24,
  2009. Proceedings}, pages 32--47, 2009.

\bibitem[BB04a]{DBLP:conf/eurocrypt/BonehB04a}
Dan Boneh and Xavier Boyen.
\newblock Efficient selective-id secure identity-based encryption without
  random oracles.
\newblock In {\em Advances in Cryptology - {EUROCRYPT} 2004, International
  Conference on the Theory and Applications of Cryptographic Techniques,
  Interlaken, Switzerland, May 2-6, 2004, Proceedings}, pages 223--238, 2004.

\bibitem[BB04b]{DBLP:conf/crypto/BonehB04}
Dan Boneh and Xavier Boyen.
\newblock Secure identity based encryption without random oracles.
\newblock In {\em Advances in Cryptology - {CRYPTO} 2004, 24th Annual
  International CryptologyConference, Santa Barbara, California, USA, August
  15-19, 2004, Proceedings}, pages 443--459, 2004.

\bibitem[BBDP01]{BellareBDP01}
Mihir Bellare, Alexandra Boldyreva, Anand Desai, and David Pointcheval.
\newblock Key-privacy in public-key encryption.
\newblock In {\em Advances in Cryptology - {ASIACRYPT} 2001, 7th International
  Conference on the Theory and Application of Cryptology and Information
  Security, Gold Coast, Australia, December 9-13, 2001, Proceedings}, pages
  566--582, 2001.

\bibitem[BF01]{BF01}
Dan Boneh and Matthew~K Franklin.
\newblock Identity-based encryption from the weil pairing.
\newblock {\em international cryptology conference}, 2001:213--229, 2001.

\bibitem[BGH07]{BGH07}
Dan Boneh, Craig Gentry, and Michael Hamburg.
\newblock Space-efficient identity based encryption without pairings.
\newblock {\em {IACR} Cryptology ePrint Archive}, 2007:177, 2007.

\bibitem[BLSV17]{BLSV17}
Zvika Brakerski, Alex Lombardi, Gil Segev, and Vinod Vaikuntanathan.
\newblock Anonymous ibe, leakage resilience and circular security from new
  assumptions.
\newblock {\em {IACR} Cryptology ePrint Archive}, 2017:967, 2017.

\bibitem[Boy03]{Boyen03}
Xavier Boyen.
\newblock Multipurpose identity-based signcryption {(A} swiss army knife for
  identity-based cryptography).
\newblock In {\em Advances in Cryptology - {CRYPTO} 2003, 23rd Annual
  International Cryptology Conference, Santa Barbara, California, USA, August
  17-21, 2003, Proceedings}, pages 383--399, 2003.

\bibitem[Boy10]{Boyen10}
Xavier Boyen.
\newblock Lattice mixing and vanishing trapdoors: {A} framework for fully
  secure short signatures and more.
\newblock In {\em Public Key Cryptography - {PKC} 2010, 13th International
  Conference on Practice and Theory in Public Key Cryptography, Paris, France,
  May 26-28, 2010. Proceedings}, pages 499--517, 2010.

\bibitem[BW06]{BW06a}
Xavier Boyen and Brent Waters.
\newblock Anonymous hierarchical identity-based encryption (without random
  oracles).
\newblock In {\em Advances in Cryptology - {CRYPTO} 2006, 26th Annual
  International Cryptology Conference, Santa Barbara, California, USA, August
  20-24, 2006, Proceedings}, pages 290--307, 2006.

\bibitem[CHKP10]{CHKP10}
David Cash, Dennis Hofheinz, Eike Kiltz, and Chris Peikert.
\newblock Bonsai trees, or how to delegate a lattice basis.
\newblock In {\em Advances in Cryptology - {EUROCRYPT} 2010, 29th Annual
  International Conference on the Theory and Applications of Cryptographic
  Techniques, French Riviera, May 30 - June 3, 2010. Proceedings}, pages
  523--552, 2010.

\bibitem[Coc01]{Cocks01}
Clifford Cocks.
\newblock An identity based encryption scheme based on quadratic residues.
\newblock In Bahram Honary, editor, {\em Cryptography and Coding}, pages
  360--363, Berlin, Heidelberg, 2001. Springer Berlin Heidelberg.

\bibitem[DG17]{DG17a}
Nico D{\"{o}}ttling and Sanjam Garg.
\newblock Identity-based encryption from the diffie-hellman assumption.
\newblock In {\em Advances in Cryptology - {CRYPTO} 2017 - 37th Annual
  International Cryptology Conference, Santa Barbara, CA, USA, August 20-24,
  2017, Proceedings, Part {I}}, pages 537--569, 2017.

\bibitem[DH76]{diffie1976new}
Whitfield Diffie and Martin~E Hellman.
\newblock New directions in cryptography.
\newblock {\em IEEE Transactions on Information Theory}, 22(6):644--654, 1976.

\bibitem[Gen06]{Gentry06}
Craig Gentry.
\newblock Practical identity-based encryption without random oracles.
\newblock In Serge Vaudenay, editor, {\em Advances in Cryptology - EUROCRYPT
  2006}, pages 445--464, Berlin, Heidelberg, 2006. Springer Berlin Heidelberg.

\bibitem[GPV08]{GPV08}
Craig Gentry, Chris Peikert, and Vinod Vaikuntanathan.
\newblock Trapdoors for hard lattices and new cryptographic constructions.
\newblock In {\em Proceedings of the 40th Annual {ACM} Symposium on Theory of
  Computing, Victoria, British Columbia, Canada, May 17-20, 2008}, pages
  197--206, 2008.

\bibitem[GSRD17]{Gupta}
Kanika Gupta, S.~Sharmila~Deva Selvi, C.~Pandu Rangan, and Shubham~Sopan Dighe.
\newblock Identity-based group encryption revisited.
\newblock 2017.

\bibitem[LRL{\etalchar{+}}16]{LuoRLHLWXW16}
Xiling Luo, Yili Ren, Jingwen Liu, Jiankun Hu, Weiran Liu, Zhen Wang, Wei Xu,
  and Qianhong Wu.
\newblock Identity-based group encryption.
\newblock In {\em Information Security and Privacy - 21st Australasian
  Conference, {ACISP} 2016, Melbourne, VIC, Australia, July 4-6, 2016,
  Proceedings, Part {II}}, pages 87--102, 2016.

\bibitem[Sha84]{Shamir84}
Adi Shamir.
\newblock Identity-based cryptosystems and signature schemes.
\newblock In {\em Advances in Cryptology, Proceedings of {CRYPTO} '84, Santa
  Barbara, California, USA, August 19-22, 1984, Proceedings}, pages 47--53,
  1984.

\bibitem[SOK00]{SOK00}
R~Sakai, K~Ohgishi, and M~Kasahara.
\newblock Cryptosystem based on pairings.
\newblock 01 2000.

\bibitem[Wat05]{Waters05}
Brent Waters.
\newblock Efficient identity-based encryption without random oracles.
\newblock In Ronald Cramer, editor, {\em Advances in Cryptology -- EUROCRYPT
  2005}, pages 114--127, Berlin, Heidelberg, 2005. Springer Berlin Heidelberg.

\bibitem[Wat09]{Waters09}
Brent Waters.
\newblock Dual system encryption: Realizing fully secure ibe and hibe under
  simple assumptions.
\newblock In Shai Halevi, editor, {\em Advances in Cryptology - CRYPTO 2009},
  pages 619--636, Berlin, Heidelberg, 2009. Springer Berlin Heidelberg.

\end{thebibliography}

\end{document}